# Photonics probing of DNA specific spatial mass density fluctuations in gut cell nuclei due to total body irradiation via confocal imaging

Mehedi Hasan,[1] Pradeep Shukla,[2] Shirsendu Nanda,[1] Prakash Adhikari,[1] Radhakrishna Rao,[2] and Prabhakar Pradhan[1]

[1]Department of Physics and Astronomy, Mississippi State University, Mississippi State, MS 39762

[2]Department of Physiology, University of Tennessee Health Science Center, Memphis, TN 38163

Contacts: pp838@msstate.edu and pshukla2@uthsc.edu

**Abstract:** Abnormalities within cells result in nanoscale structural alterations can be characterized via confocal imaging and quantification of these alterations. Accidental or deliberate exposure to total body irradiation (TBI) have adverse effects on the nuclear DNAs of cells. Here, we study the DNA molecular mass density spatial structural alterations of chromatin in cell nuclei of gut tissues caused by the exposure to standard doses of 4Gy TBI, using the light localization technique called inverse participation ratio (IPR) in confocal images. Results indicate radiation suppresses DNA spatial mass density fluctuations. And hence, reduction and saturation in DNA density fluctuations are observed on different durations of post-irradiation.

## 1. Introduction

It is now reported that abnormalities in a cell are associated with the structural alterations of the basic building blocks of a cell at the nanoscale, such as DNA, RNA, protein, etc. The structural changes can happen at the molecular specific spatial mass density changes to total mass density changes in a cell. The structural alternations at the nanoscale level in cells provide a plethora of information that can help us to predict the physiological state of cells. The structural changes in cells/tissues can be quantified using light localization techniques in confocal images at the nano to submicron scale levels (1,2). Most often the prominent structural changes happen in the cell in the DNA/chromatin, which is highly susceptible to damage, in the abnormalities. These structural abnormalities in cells could be due to diseases, radiation exposure from sources such as ultraviolet light, mutagenic chemicals, heavy reactive processes, and ionizing radiation (IR)(3,4). In this paper, we will study the abnormalities in gut cells due to exposure to TBI.

Radiation therapy is commonly used as a method of treatment for different types of diseases including the treatment of different cancer types. One can be exposed to radiation accidentally or by radiation treatment. Accidental total body radiation exposition happened due to environmental hazards in radiation disasters. In radiation treatment or therapy, normal cells/tissues are unintentionally exposed to radiation. Hence the structural alterations are observed in the cell nuclei due to total body irradiation. These alterations vary as time progresses and intensities of radiation. At this point it was observed that in each proceeding hour post-ionized radiation, different structural effects are detected within the cell nuclei (5). The tissues of the organs that are exposed to the radiation play a central role in determining the radiation-tolerance levels of those exposed tissues. The radiation effects on the normal cells could have long-term effects on the patients, due to the radiation related damage. The patients may experience the radiation damage symptoms soon after the radiation therapy or the symptoms may take up to years to appear (6). However, the pathological processes of radiation damage in cells begin soon after the normal tissues are exposed to radiation.

Damages due to radiation can be divided into two categories based on the amount of time it takes for the symptoms to become noticeable. They are acute and consequential. Acute effects are noticeable within a few weeks after the radiation therapy treatment; sometimes even within this period. However, this damage is at the nano-scale levels and hardly distinguishable by conventional imaging techniques. On the other hand, consequential effects are prominent in the later stages of radiation exposure with severe health issues occurring. These two categories of radiation effects are initially at the nanoscale level in cells/tissues and have not been studied or understood well before. In reality, these processes are very complex, and the radiation can trigger a range of phenomena that may go unnoticed. In some cases, late or consequential effects have been reported up to 34 years after the radiation exposure(7,8).

Organ damage is one of the many side-effects that is noted during the post irradiation period because of TBI. When an organ is exposed to radiation, its tissues may react in different ways based on the radiation dose and their level of radiation tolerance. Even though the exact level of tolerance of an organ is unknown, the estimated tolerance doses are reported in published guidelines. Radiation results in an ionization event and the production of free radicals. Further, it is known that TBI causes severe damage to different cellular components such as cytoplasm, chromatin, etc. in a cell. When cells are in their first cell division or the process of their first cell division after being exposed to radiation, their DNA molecules are likely to get damaged partially to severely, if the cells are exposed to the radiation (9,10). Chromosomal damage that is either unrepaired or improperly repaired causes mitosis or mitotic death (11). Double-strand break (DSB) damage due to radiation causes the DNA double helix to break in such a way that it becomes nearly impossible to keep the two broken ends at the same place. In addition, as there is less chance of it repairing itself, this damage can even cause unsuitable recombination of the DNA genome. Such inappropriate repair of a DSB can lead to massive instability in the genome that can run through generations of chromosomal fragments (12). One single DSB event can even cause apoptosis, i.e. the death of the whole cell (13). Consequently, this DNA damage carries long-term effects such as genetic instability (14).

During the post total body γ-irradiation period, the intestinal tissues of mice can experience adverse effects. This includes but not limited to the induction of oxidative stress and apoptosis in the intestinal tissues (15). The effects of radiation also include activation of cellular signaling pathways, which ultimately leads to the expression and activation of proinflammatory and profibrotic cytokines, vascular injury, and coagulation cascade(16,17). Exposition to radiation can have a more damaging effect on patients with pre-existing conditions/diseases. As radiation causes damage that hinders the restoration of DNA, patients with ataxia-telangiectasia (a type of genetic abnormality) will experience serious radiation reactions (18). Other genetic factors can also play a role in radiosensitivity as suggested by studies on various strains of mice (19). However, it is not completely reliable to depend on the experiments of radiosensitivity on cells that were isolated from the patient except when there is a case of extremity. No significant late damages have been found in patients who showed early responses to radiation effects (20).

Treatment of irradiated tissues is difficult or sometimes impossible, due to its damage to the DNA level. Different studies and technologies have developed to increase the efficacy of conventional radiotherapy methodologies to reduce the effect of radiation in adjacent cells/tissues, or proximal damage in radiation. The 3-dimensional conformal radiotherapy (3DCRT) and intensity-modulated radiotherapy (IMRT) are two such examples of imaging and computer technology that can distribute necessary doses to cancer or tumor cells and avoid unnecessary exposure of normal tissues by proximal damage (21). Also, different forms of chemical treatment can be used before the radiation to protect normal tissues from acute as well as late radiation damages (22).

In this paper, we report a quantitative analysis of the DNA molecular mass density spatial structural changes in gut cell nuclei due to the exposer to TBI. It is now known that the gut is an organ that gets damaged heavily in irradiation. The DNA molecular mass density fluctuations are probed by using a mesoscopic physics-based spectroscopic technique, the inverse participation ratio (IPR), on DAPI stained DNA molecular (i.e. chromatin) confocal images. The molecular specific light localization properties of mass density variation due to radiation exposure are quantified as the measures of the degree of structural disorder, $L_d$, of DNA, and compared. We show that these structural changes consequently trigger the alteration of structural disorder strength, $L_d$, at the nano- to sub-micron scales cellular level of tissue. The IPR technique is a powerful tool, that has been earlier used to probe the nanoscale structural abnormalities in cells using a confocal image to distinguish stages of cancer abnormalities as well as drug-effects in abnormal cells(23,24).

## 2. Method
### 2.1. Sample Preparation:

All the experiments on animals strictly followed the guidelines provided by the Institutional Animal Care and Use Committee (IACUC) at the University of Tennessee Health Science Center. Gut tissue samples from mice were housed in an institutional animal care facility. The animal care facility replicated the regular living atmosphere of the animals by providing 12:12-hour light-dark cycles and access to regular laboratory food and water until the experiments were conducted. For all our experiments, C57BL/6 mice were used (age 12–14 weeks, collected from Harlan Laboratories, Houston, TX). FITC-inulin (50 mg/ml solution, 2 l/g body weight) was directly injected into the tail vein of the mice. In our first experiment, a Mark I, model 25, 137Cs source irradiator (JL Shepherd and Associates, San Fernando, CA) was used to perform TBI. The total irradiation dose was 4 Gy at a dose rate of ~76 cGy/min on the 12-to 14-week-old adult mice. The mice were euthanized at different hours (2, 6, 8, and 24 hrs) after the completion of the TBI dose. The mice colons are removed and cryofixed.

### 2.2. Confocal microscopy imaging:

For our confocal microscopy imaging, colon cryosections of thickness 10 and 12 microns respectively were set in an acetone-methanol mixture (1:1 ratio) for two minutes (temperature -20°C). Rehydration was performed on the sections using phosphate-buffered saline (PBS). 0.5% Triton X-100 was used to permeabilize the sections in saline for 15 minutes. The tissue sections were permeabilized with 0.2% Triton X-100 in PBS for 10 min and blocked in 4% nonfat milk in Triton-Tris buffer (150 mM sodium chloride containing 10% Tween 20mM and 20mM Tris, pH 7.4). It was then incubated for 1 hour with the DAPI, followed by incubation for 1 hour with secondary antibodies Cy3-conjugated anti-rabbit IgG antibodies and 10 min incubation with Hoechst 33342. A Zeiss 710 confocal microscope was used to examine the fluorescence and to collect the confocal imaging data. We have collected the x-y images (size 1 micron) using ZEN (Zeiss Efficient Navigation) software. These images were stacked using ImageJ software (National Institutes of Health, Bethesda, MD) and further processed using Adobe Photoshop Software (Adobe Systems, San Jose, CA).

**2.3 *The Inverse Participation Ratio (IPR) technique for the structural disorder analysis from confocal imaging or micrographs***

The mesoscopic physics-based molecular specific imaging method, Inverse Participation Ratio (IPR) using confocal imaging or "Confocal-IPR" technique, has been proven to be useful in quantifying the structural molecular changes in biological cells(25,26). The IPR technique can be applied to quantify the structural changes at the nano to submicron scales in cells. The technique is used to evaluate the localization properties of the optical lattice that are formed from the molecular specific spatial mass density distribution by using confocal imaging. Based on the light localization strength of the medium, the degree of structural disorder strength is calculated. The average and standard deviation of the IPR are proportional to the mass density fluctuations of weakly optical disordered media such as cells. The details of the method are described in Ref.(27,28); in the following, we describe the method in short for the completeness of this paper.

Consider $dV=dxdydz$ is a small finite volume of the cell slice at a point (x, y) with thickness $dz$, and $\rho$ is the DNA molecular mass density in the voxel $dV$ of the sample. In the confocal image, if $I(r)$ is the pixel intensity at position $r$, then $I(r)$ can be denoted as $I(r) \propto dV(\rho)$ (1). The local refractive index of the cell slice at a point *(x,y) i.e. n(x,y)* is directly proportional to the local mass density of the cell as $n(x,y) = n_0 + dn(x,y)$. Here, $n_0$ is the average refractive index of the confocal images, and *dn(x,y)* is the refractive index fluctuations of the voxel *dV* at position *(x,y)* (26). Here the refractive index fluctuations are less than the average refractive index $n_0$ *(dn<<$n_0$)*. We can represent the pixel intensity *I(x, y)* of the confocal image at position *(x, y)* which is linearly proportional to the refractive index *n(x, y)* of the voxel(27,28), i.e.:

$$n_0 + dn(x,y) \propto I_0 + dI(x,y)$$

If $\varepsilon_i(x,y)$ represents the optical potential corresponding to the pixel position *(x, y)* of the two-dimensional plane of the confocal image. Then the optical potential of the voxel point *(x, y)* is calculated to generate an optical lattice and represented as (1):

$$\varepsilon_i(x,y) = dn(x,y)/n_0.$$

Tight Binding Model (TBM) is commonly used to calculate the disorder properties of electrical and optical systems. The spatial structural disorder strength of an optical lattice can be analyzed by the Hamiltonian approach of the Anderson Tight Binding Model (TBM), which can be written as(29–31):

$$H = \sum_i \varepsilon_i |i><i| + t \sum_{\langle ij \rangle} (|i><j| + |j><i|). \tag{1}$$

Here, $\varepsilon_i(x,y)$ is the optical potential energy of the *i-th* lattice site, *|i>* and *|j>* are the eigenvectors of the *i-th*, and the *j-th* lattice sites and *t* is the overlap integral between sites *i* and *j*. The average IPR value, i.e. *<IPR>* of the entire sample images, can be calculated using the eigenfunctions ($E_i$'s)(27,28,32),

$$<IPR>_N = \frac{1}{N} \sum_{i=1}^{N} \int_0^L \int_0^L E_i^4(x,y) dx dy. \tag{2}$$

Here, *N* $(=(L/dx)^2, dx=dy)$ is the total number of lattice points on the refractive index matrix, and $E_i$ is the *i-th* eigenfunction of the Hamiltonian H. It is shown that the calculated *<IPR>=<<IPR>$_N$>$_{ensemble}$* is directly proportional to the degree of structural disorder strength represented by $L_d$. For Gaussian white noise potential, $L_d = <dn> \times l_c$, where *<dn>* is the average refractive index fluctuations and $l_c$ is the spatial correlation length of the refractive index fluctuations over the sample (1,2). Therefore,

$$\langle IPR \rangle \propto \sigma(<IPR>) \propto L_d = <dn> \times l_c. \tag{3}$$

Further statistical analysis is performed by computing the average and standard deviation of the DNA molecular specific structural alteration or the IPR values of the confocal images.

## 3. Results and Discussions

In this experiment, mice were exposed to TBI to evaluate the effects of ionizing gamma radiation (IR) on DNA molecular spatial mass density of nuclei from gut cells. For this, mice were fed a normal diet and were divided into 5 groups. Each group of mice was irradiated in a radiation chamber of total body irradiation (TBI). The mice were then sacrificed after post-irradiation (postIR) time points: 2, 6, 8, and 24 hours, and then their guts were removed and cryo-preserved. Consequently, the gut tissues were prepared for DAPI staining of cell nuclei as described in the Method section. In particular, by using confocal microscopy, the images of DAPI stained tissues were taken and 4Gy ionizing gamma radiation treated gut tissues are defined as follows: Sham (control), 2hr_postIR, 6hr-postIR, 8hr_postIR, 24hr_postIR. In the next step, the confocal micrographs are analyzed by the IPR technique that is described in the method for the quantification of the DNA molecular density structural disorder.

In Fig.1 (a)-(e), representative confocal images of DAPI stained nuclei of the mice gut tissue samples are presented for sham (control) and total body irradiated mice for different post-irradiation durations: (a) Sham (control), (b) 2 hours post-irradiation, (c) 6 hours post-irradiation, (d) 8 hours post-irradiation, and (e) 24 hours post-irradiation; and their corresponding <IPR> images are shown in figures Figs.1. (a'), (b'), (c'), (d'), and (e'), respectively. The <IPR> images represent the structural abnormalities of the chromatin in the cell nuclei of gut tissue samples. In the IPR images, red color represents higher mass density fluctuations within the pixel of IPR images. On the other hand, the blue color represents the lower mass density fluctuation for every pixel of IPR images. The bar graphs of the std(<*IPR*>) or $\sigma(<IPR>)$ quantified for the ensemble averaged DNA mass density fluctuations using the confocal-IPR technique for gamma-irradiated mice cell nuclei from gut tissues are presented in Fig. 2.

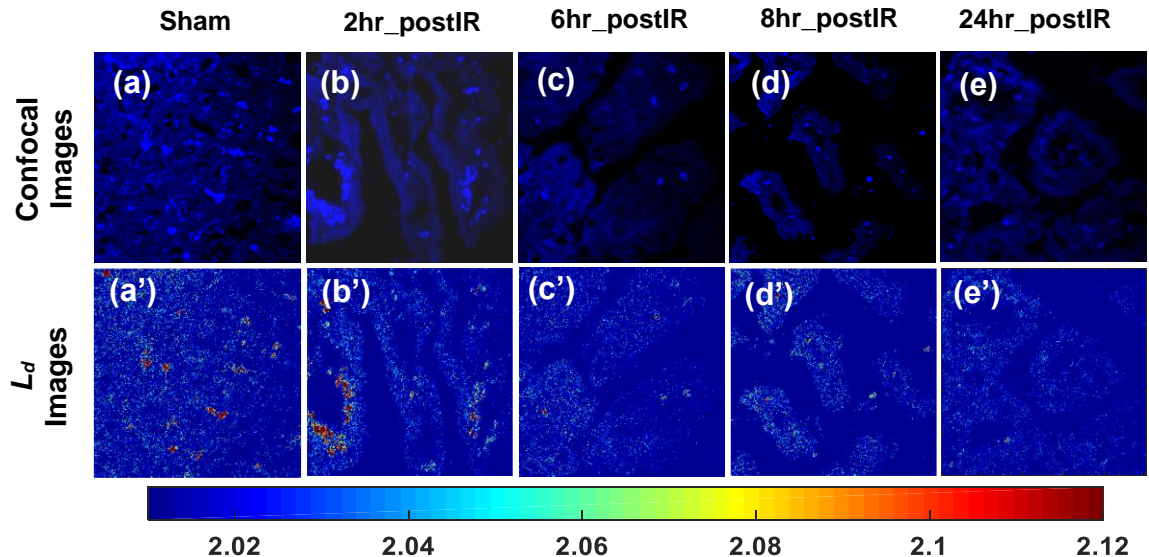

**Fig.1.** (a)-(e) represent the confocal images of gut tissues stained with DAPI for: (a) sham (control), (b) 2hr_postIR, (c) 6hr_postIR, (d) 8hr_postIR, and (e) 24hr_postIR, and their corresponding IPR images are shown in (a'), (b'), (c'), (d'), and (e'), respectively. The IPR images show the structural disorder properties in gut tissue cell nuclei and the effect of ionizing gamma radiation (4Gy) on their DNA molecules/chromatins.

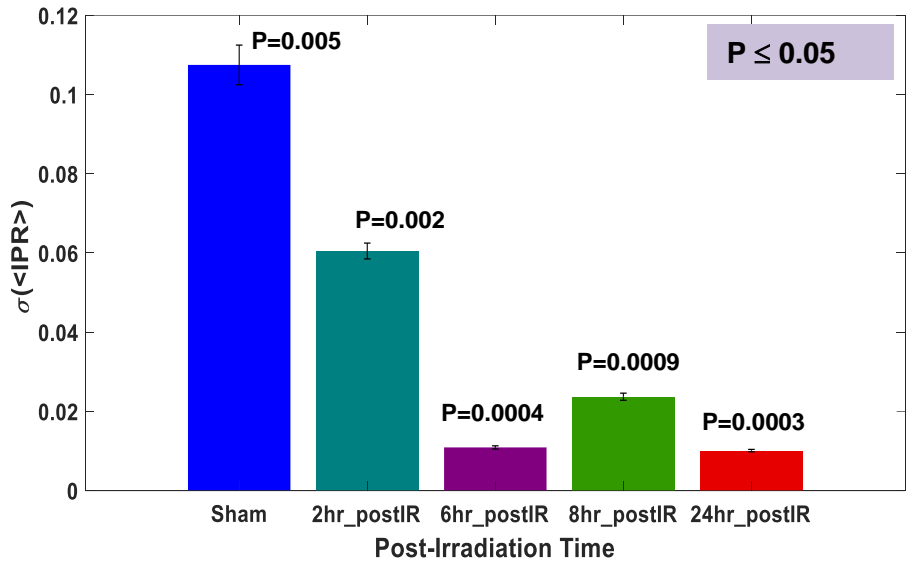

**Fig.2.** Bar graphs of the ensemble averaged std(<*IPR*>) of DAPI stained mice gut tissue nuclei for: (a') Sham (control), (b') 2 hours post-irradiation, (c') 6 hours post-irradiation, (d') 8 hours post-irradiation, and (e') 24 hours post-irradiation, based on Figs. (a')-(e'). The IPR analysis illustrates that the $\sigma(<IPR>)$ or the degree of disorder strength ($L_d$) of the ionizing gamma radiation treated mice gut tissues decreases by 43.72% after 2 hours post-irradiation, 89.86% after 6 hours post-irradiation, 77.95% after 8 hours post-irradiation and 90.60% after 24 hours post-irradiation relative to the sham (control). The p-values for each of the postIR groups are statistically significance i.e. $\leq 0.05$ compared to the sham.

These bar graphs in Fig. 2 show a significant decrease in the $\sigma(<IPR>)$ or DNA mass densities of mice cell nuclei with the increasing elapsed duration after 4Gy ionizing gamma radiation exposure compared to the sham (control). From Fig. 2, the statistical analysis shows that the DNA molecular structural disorder strength ($L_d$) of ionizing gamma irradiation treated DNA nuclei chromatin structure decreases as follows: 43.72% after 2 hours post-irradiation (2hr_postIR), 89.86% after 6 hours post-irradiation, 77.95% after 8 hours post-irradiation, and 90.60% after 24 hours post-irradiation compared to the control (Sham). The results imply that after the exposure to the gamma radiation, the DNA spatial mass density fluctuations were decreased, in general. The cause of this suppression is the decrease in the DNA mass density fluctuations with the increasing elapsed duration after radiation exposure. Therefore, the continuous decrease in the degree of disorder strength of 4Gy gamma-irradiated mice gut tissue DNA structure after different durations of postIR hours suggests that TBI has adverse effects on cell nuclei at the nano- to submicron-scale levels.

The results illustrate that the effects of 4Gy gamma radiation effect on the chromatin in cell nuclei of gut tissues at the nanoscale level may vary with time and worsen with increasing elapsed hours after exposure and eventually the fluctuations saturates to a lower value. Results also indicate that the nuclear changes by irradiation are very quick, similar to its effect on bone marrow. The effect is observable as quick as 2-hr post-irradiation and reaches the maximum by 24- hr post-irradiation. The time course correlates with functional changes such as tight junction disruption (5). The degree of mass density fluctuations acts as a potential biomarker for measuring the nano to submicron scales spatial structural alteration of DNA mass density fluctuations in the nuclei of the cells of gut tissues. In particular, the molecular specific IPR analysis shows that the degree of structural disorder strength $(L_d)$ at the nanoscale level in cell nuclei chromatin is reduced with the exposure to ionizing gamma radiation or TBI. The change in the degree of DNA molecular spatial structural changes are measured based on their light localization strength. The $\sigma(<IPR>)$ value is calculated using confocal images which are directly proportional to the degree of structural disorder strength as mentioned in the Methods section. To understand the molecular specific structural properties of irradiated cells using the mesoscopic physics-based imaging technique, the degree of the structural disorder strength is calculated in comparison to the sham (control).

## 4. Conclusions

In this work, a recently developed confocal-IPR technique has been used to report the effect of radiation on DNA molecular specific mass density fluctuations in the nuclei of gut tissues. Here, the IPR value of confocal images of control and gamma radiation (4Gy) affected nuclei of cells from mice gut tissues are calculated which is directly proportional to the degree of structural disorder strength. These statistical measures of $<IPR>$ acts as a potential biomarker for the measurement of the alteration of the DNA molecular mass density fluctuations. It has been reported that irradiation effects vary with the post-irradiation duration time and have adverse effects on the DNA spatial structural arrangement. However, there was a small abnormality in the gut tissues' DNA structure after 8 hours post-irradiation where the $\sigma(<IPR>)$ has a higher value than 6-hours as well as 24-hours post-irradiations. This increase in the $\sigma(<IPR>)$ value may be due to the irradiated tissues trying to retain their initial structural properties, and this requires more experimental investigations.

Finally, the decrease in the structural disorder strength in this study confirms that irradiation affects the DNA nuclei chromatin structure at the nanoscale level in mice gut tissues. In particular, the radiation suppresses the DNA mass density fluctuations and eventually it gets saturated with the increase of the post-irradiation time. Suppression of the DNA mass density fluctuations can affect many activities in the chromatin including the daily nuclear transcriptions, in turn, the DNA replication which may result in genetic alterations as well. In addition, the nanoscale study of molecular specific irradiation affected cells explains their physical states and helps us to improve radiation therapy methodologies which inevitably involve irradiation of normal cells.

**Funding**

The part of this work is supported by MSU to PPradhan and the NIH DK55532 to RKRao

**Acknowledgments**

Funding from MSU and NIH are acknowledged.

**Disclosers**

The authors have no conflict of interest.